\documentclass[letter]{aa}  
\usepackage{amsmath}
\usepackage{amssymb}
\usepackage{natbib}
\usepackage{graphicx}
\usepackage{txfonts}
\usepackage[english]{babel}
\usepackage{hyperref}
\usepackage{units}
\begin{document}

\title{Connecting the shadows: probing inner disk geometries using shadows in transitional disks}

\author{
        M. Min\inst{1,2}
                \and
        T. Stolker\inst{2}
                \and
        C. Dominik\inst{2}
                \and
        M. Benisty\inst{3,4}
}

\offprints{M. Min, \email{M.Min@sron.nl}}

\institute{
SRON Netherlands Institute for Space Research, Sorbonnelaan 2, 3584 CA Utrecht, The Netherlands
        \and
Anton Pannekoek Institute for Astronomy, University of Amsterdam, Science Park 904, 1098 XH, Amsterdam, The Netherlands
        \and
Unidad Mixta Internacional Franco-Chilena de Astronom\'{i}a, CNRS/INSU UMI 3386 and Departamento de Astronom\'{i}a, Universidad de Chile, Casilla 36-D, Santiago, Chile
        \and
University Grenoble Alpes, CNRS, IPAG, 38000 Grenoble, France
}

   \date{Date: \today}

 
  \abstract
   {}
   {Shadows in transitional disks are generally interpreted as signs of a misaligned inner disk. This disk is usually beyond the reach of current day high contrast imaging facilities. However, the location and morphology of the shadow features allow us to reconstruct the inner disk geometry.}
   {We derive analytic equations of the locations of the shadow features as a function of the orientation of the inner and outer disk and the height of the outer disk wall. In contrast to previous claims in the literature, we show that the position angle of the line connecting the shadows cannot be directly related to the position angle of the inner disk. }
   {We show how the analytic framework derived here can be applied to transitional disks with shadow features. We use estimates of the outer disk height to put constraints on the inner disk orientation. In contrast with the results from \citet{2017ApJ...838...62L}, we derive that for the disk surrounding HD~100453 the analytic estimates and interferometric observations result in a consistent picture of the orientation of the inner disk.}
   {The elegant consistency in our analytic framework between observation and theory strongly support both the interpretation of the shadow features as coming from a misaligned inner disk as well as the diagnostic value of near infrared interferometry for inner disk geometry.}


   \maketitle
%

\section{Introduction}

The advent of high spatial resolution imaging is revolutionising our
understanding of the structure of protoplanetary disks.  Scattered
light imaging provides a view on disks that is focused on the location
and properties of the disk surface.  It gives direct access to
the structure of the disk surface by tracing smaller grains embedded
in the disk gas, allowing to identify gaps, spirals and other
structures in the disk, as well as determining dust properties 
\citep[see e.g.][for some examples of the rapidly growing literature]{2012ApJ...748L..22M, 2013ApJ...762...48G, 2014ApJ...790...56A, 2016A&A...595A.113S, 2017A&A...599A.108M}. 
On the other hand, in contrast to thermal imaging (e.g. with ALMA),
scattered light imaging relies on light from the central light source,
the star, reaching the disk surface directly, making the disk
appearance sensitive to shadowing and light-travel time effects
\citep{2016A&A...593L..20K}.  This makes scattered light imaging
sensitive to the structure and variability of the disk structure
closer to the star than what can be reached by even the best
high-contrast imaging technology available today \citep[see e.g.][for
an interesting case of possibly variable
shadows]{2016A&A...595A.113S}.  In this way, studying shadows in the
outer disk can form a bridge between outer disk imaging and
interferometric observations of the hottest, innermost parts of the
disk.

Recent observations have led to the stunning revelation that the inner
parts (by ``inner'' we here refer to a part of the disk that has not
yet been resolved in the observations) of protoplanetary disks can not
only display small warps \citep[as suggested by][to explain the shadowing features in the disk surrounding TW~Hya]{2017ApJ...835..205D}, 
but can be strongly misaligned with respect
to the outer disk.  In at least two sources, an inner disk structure
appears to be misaligned so strongly that two clear shadow lanes
appear on the outer disk \citep{2015ApJ...798L..44M,
  2017A&A...597A..42B}.  This can happen if the inclination of the
inner disk relative to the outer disk is larger than the sum of the
opening angle (aspect ratios) of the outer and inner disks
\citep{2015ApJ...798L..44M,2017ApJ...838...62L,2016A&A...595A.113S}. The
thickness of the shadow lanes allows conclusions about the vertical
structure of the shadow-casting inner disk.
Misaligned, very compact inner disks could be due to the interaction with the magnetic
field of the star, as in the classical case of AA Tau
\citep{1999A&A...349..619B}, or might be caused by the interaction of
a massive companion on an inclined orbit embedded in the disk.  The
latter would be especially interesting in the context of the measured
relative inclinations between the stellar rotation axis and orbit
orientation in a number of exoplanetary systems \citep[see e.g. Fig.~7
in][]{2014ApJ...784...66X}, indicating that these differences can
originate early in the history of a planetary system - that is, still in
the formation/disk phase.

Connecting the location of the two shadow lanes to the relative
orientation of the inner and outer disks sometimes defies simple
intuition, which is why we address this problem here in a general
way. The interpretation is complicated by the fact that the inner and
outer disks are not perfectly flat and by the fact that the system is
not seen face-on.  These aspects must be properly taken into account when deriving the geometric parameters of the inner disk.

In Section \ref{sec:equations} of this letter, we derive analytical
formulas for the location of the shadow lanes cast by a highly
misaligned inner disk onto the surface of a geometrically thick outer
disk. In Section \ref{sec:HD100453} we demonstrate how the equations
can be applied to the shadows seen in HD~100453 and compare the
derived geometry with values obtained for the innermost hot disk
regions through near-IR interferometry. In particular we show that, in
contrast to claims by \cite{2017ApJ...838...62L}, the geometric
parameters derived for HD~100453 in \cite{2017A&A...597A..42B}, are
correct.

\section{Analytic equations for shadow locations}
\label{sec:equations}

We define the coordinate system where we put the positive $x-$axis towards the North, the positive $y-$axis towards the East, and the positive $z-$axis towards the observer.
The normal vector of the inner disk, $\hat{n}_1$, and outer disk, $\hat{n}_2$ have to be rotated according to their respective inclination and position angles, $\theta_{1,2}$ and $\phi_{1,2}$. We rotate the vectors along the $x$-axis for inclination with the rotation matrix
\begin{equation}
R_\theta = \begin{bmatrix}
        1       & 0                     & 0                     \\[0.3em]
        0       & \cos(\theta)  & \sin(\theta)  \\[0.3em]
        0       & -\sin(\theta) & \cos(\theta)
     \end{bmatrix},
\end{equation}
and after that for the position angle along the $z$-axis with the matrix
\begin{equation}
R_\phi = \begin{bmatrix}
        \cos(\phi)      & -\sin(\phi)   & 0     \\[0.3em]
        \sin(\phi)      & \cos(\phi)    & 0     \\[0.3em]
        0               & 0                     & 1
     \end{bmatrix}.
\end{equation}
This gives for the normal vectors of the rotated disks:
\begin{equation}
\hat{n} = \begin{bmatrix}
        -\sin(\theta)\sin(\phi) \\[0.3em]
        \sin(\theta)\cos(\phi)  \\[0.3em]
        \cos(\theta)
     \end{bmatrix}.
\end{equation}
These two normal vectors define the planes of the two disks. We assume for the remainder that the inner disk is very thin, and the shadows are cast on the scattering surface of the outer disk which is lifted a distance $h$ above the midplane. We note that the scattering surface is the height in the disk where the radiation hits an optical depth $\tau=1$. This is not to be confused with the scale height of the disk, $H$, which is usually significantly lower.

The misalignment of the disks, $\Delta\theta$, is simply given by the angle between the normal vectors
\begin{eqnarray}
\Delta\theta&=&\cos^{-1}\left[\hat{n}_1\cdot\hat{n}_2\right] \\
&=&\cos^{-1}\left[\sin(\theta_1)\sin(\theta_2)\cos(\phi_1-\phi_2)+\cos(\theta_1)\cos(\theta_2)\right]. \nonumber 
\end{eqnarray}
The angle between the shadows as measured in the plane of the outer disk is
\begin{equation}
\omega=2\tan^{-1}\sqrt{\frac{\tan^2(\Delta\theta)}{(h/R)^2}-1}.
\end{equation}

\begin{figure}[!tp]
\centerline{\resizebox{0.9\hsize}{!}{\includegraphics{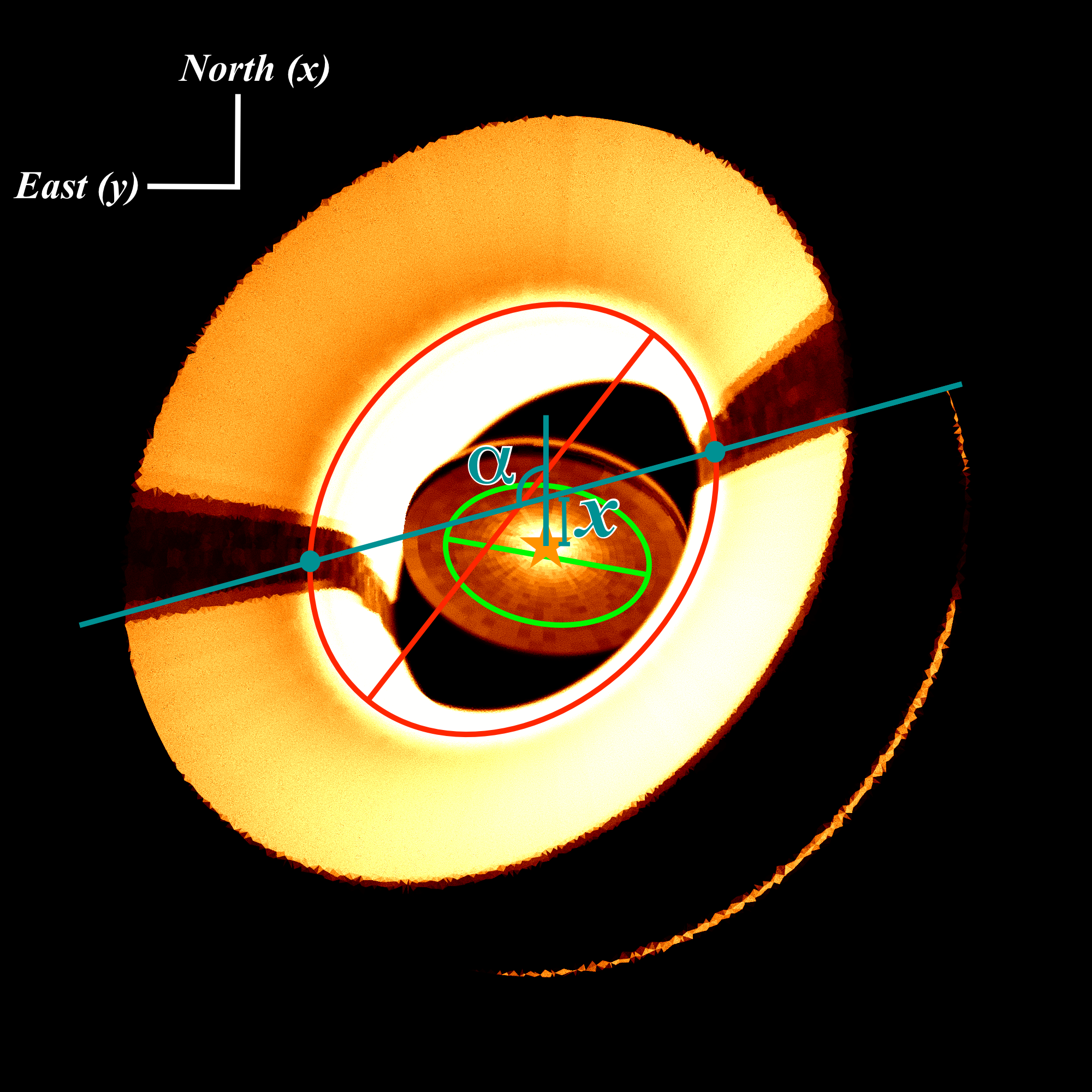}}}
\caption{Cartoon representation of the shadows cast on the outer disk of a transitional disk by a misaligned inner disk. For the purpose of clarity, the inner disk is blown up significantly to be able to better visualise the geometry. Indicated are the ellipses of the inner and outer disks showing their position angles. Also indicated by the blue line is the connecting line between the shadows. Here $\alpha$ is the position angle of this line (see also Eq.~\ref{eq:posangle}) and $x$ the offset towards the North (see also Eq.~\ref{eq:offset})}
\label{fig:cartoon}
\end{figure}

\begin{figure*}[!htbp]
\begin{center}
\resizebox{\hsize}{!}{\begin{tabular}{cccc}
$\theta_1=45^\circ, \theta_2=-38^\circ$, $\phi_1=82^\circ, \phi_2=142^\circ$ & 
$\theta_1=45^\circ, \theta_2=38^\circ$, $\phi_1=82^\circ, \phi_2=142^\circ$ & 
$\theta_1=0^\circ, \theta_2=-38^\circ$, $\phi_1=82^\circ, \phi_2=142^\circ$ & 
$\theta_1=45^\circ, \theta_2=-38^\circ$, $\phi_1=108^\circ, \phi_2=142^\circ$ \\
\resizebox{0.33\hsize}{!}{\includegraphics{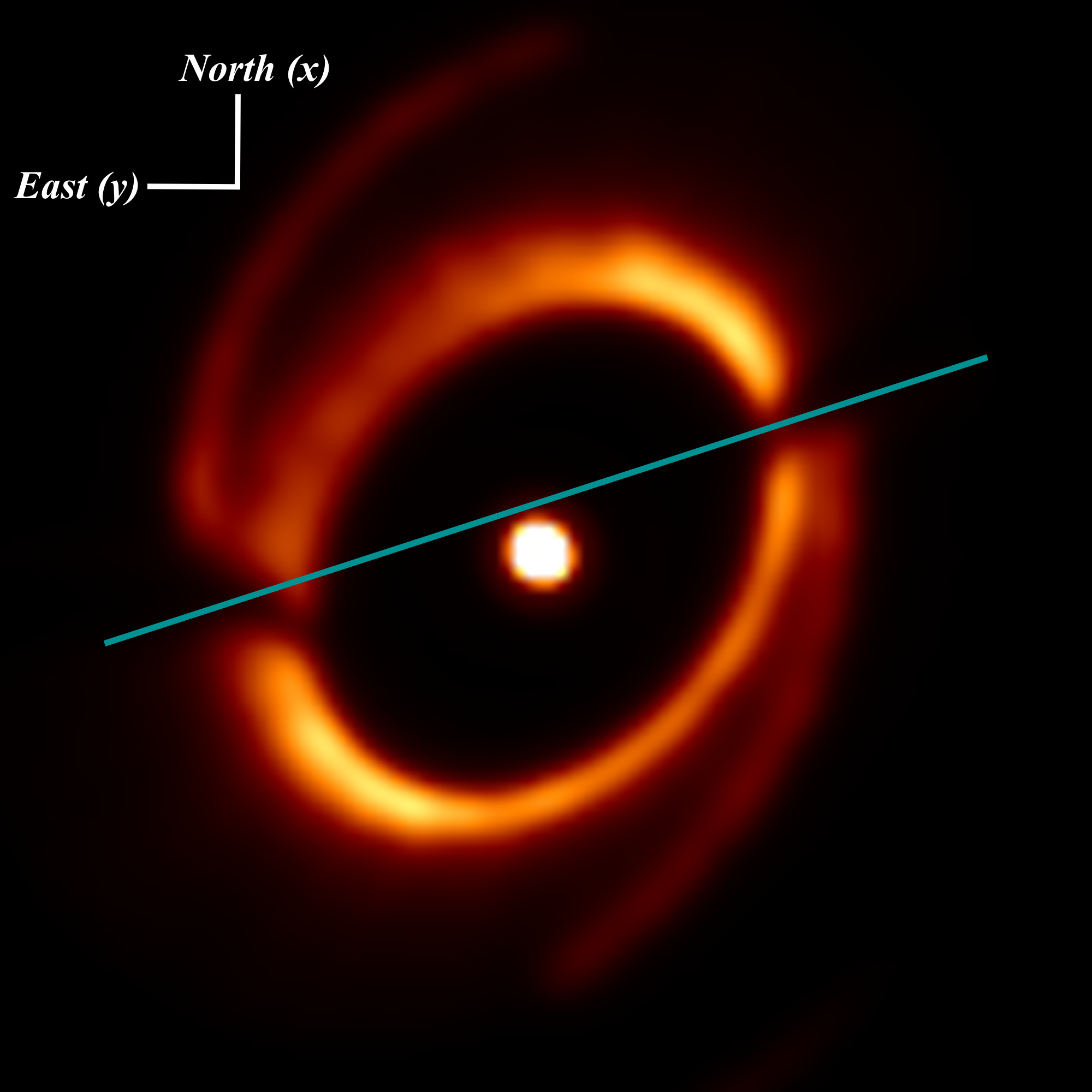}} &
\resizebox{0.33\hsize}{!}{\includegraphics{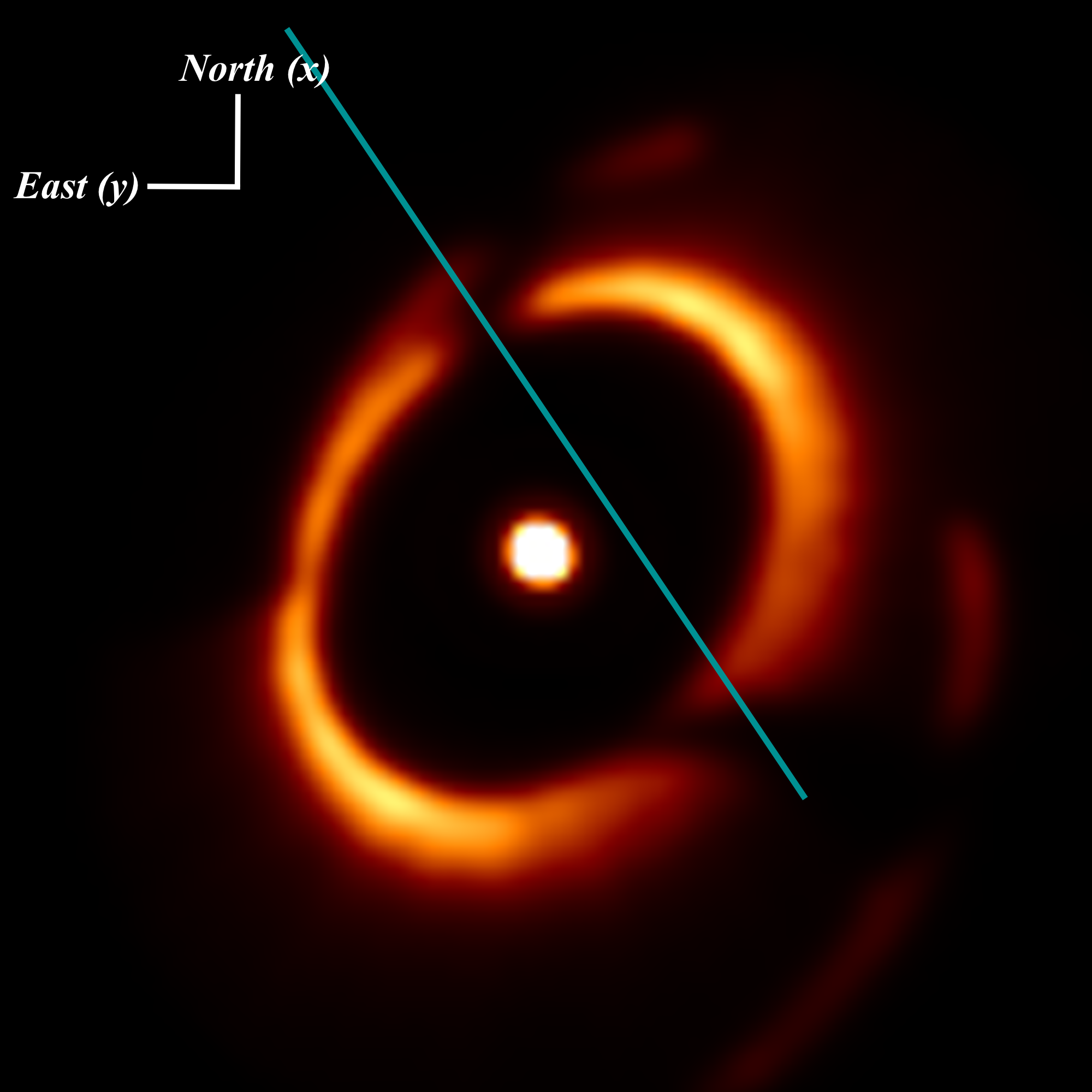}} &
\resizebox{0.33\hsize}{!}{\includegraphics{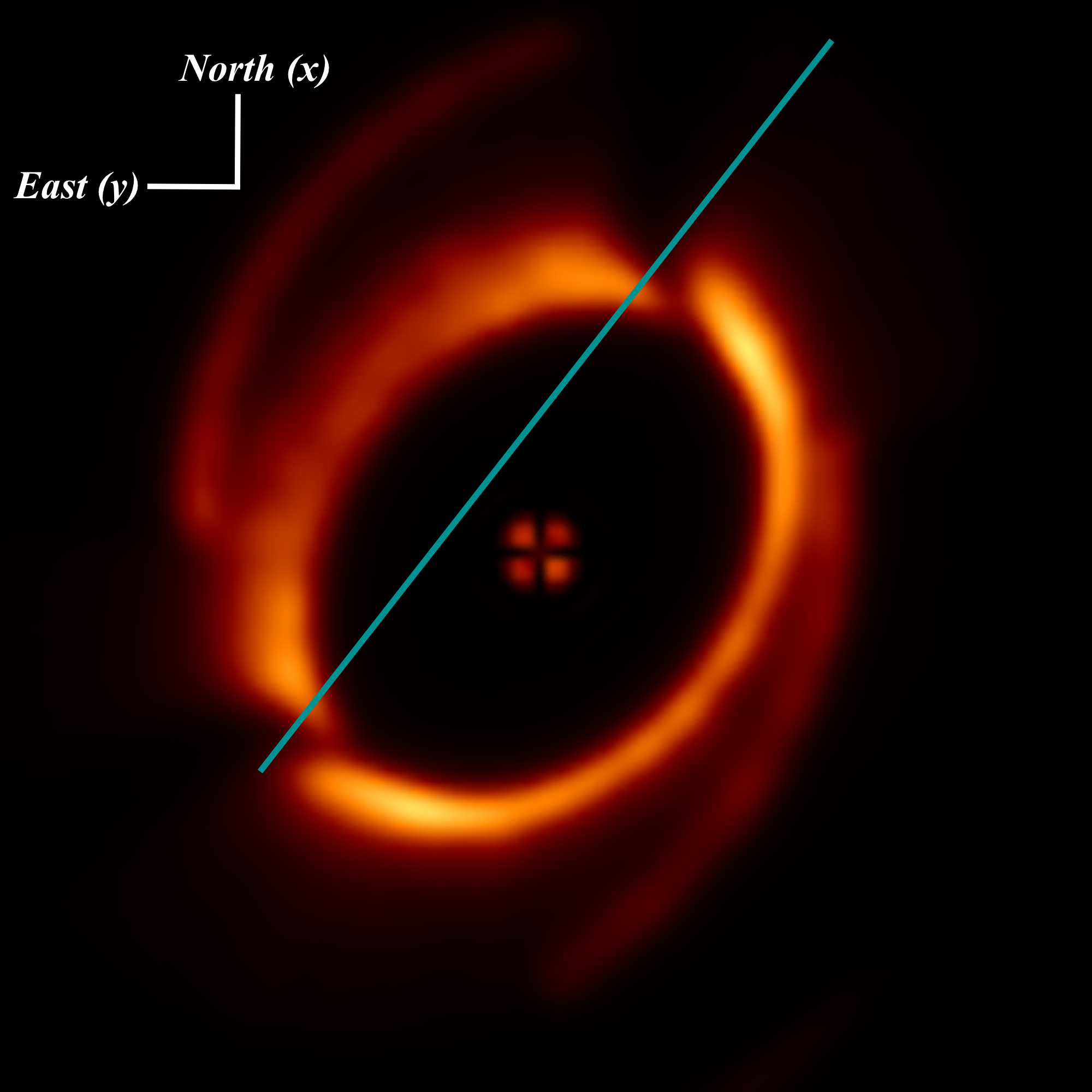}} &
\resizebox{0.33\hsize}{!}{\includegraphics{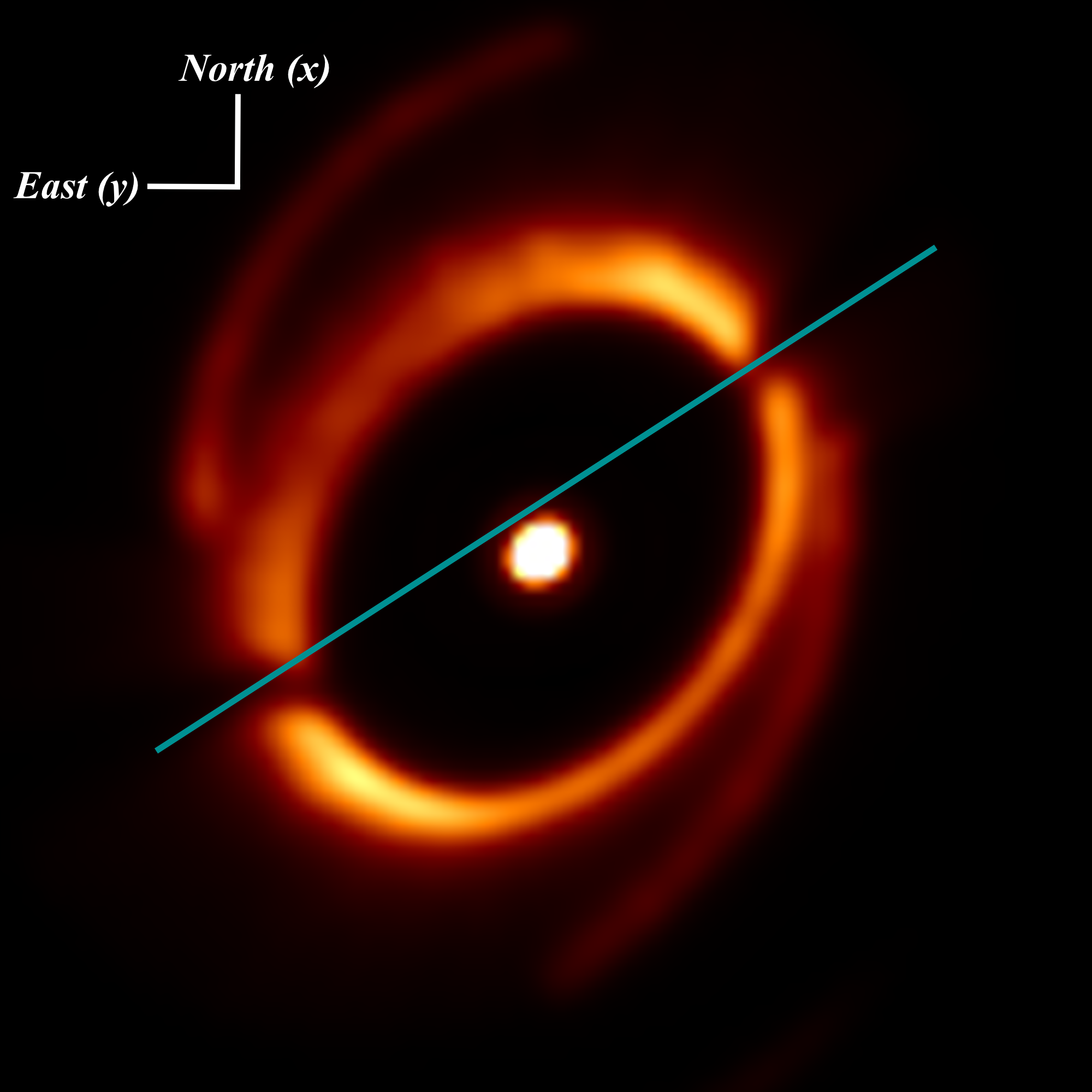}} \\
$\theta_1=90^\circ, \theta_2=-38^\circ$, $\phi_1=82^\circ, \phi_2=142^\circ$ & 
$\theta_1=-45^\circ, \theta_2=0^\circ$, $\phi_1=82^\circ, \phi_2=142^\circ$ & 
$\theta_1=45^\circ, \theta_2=0^\circ$, $\phi_1=82^\circ, \phi_2=142^\circ$ & 
$\theta_1=-45^\circ, \theta_2=38^\circ$, $\phi_1=82^\circ, \phi_2=142^\circ$ \\
\resizebox{0.33\hsize}{!}{\includegraphics{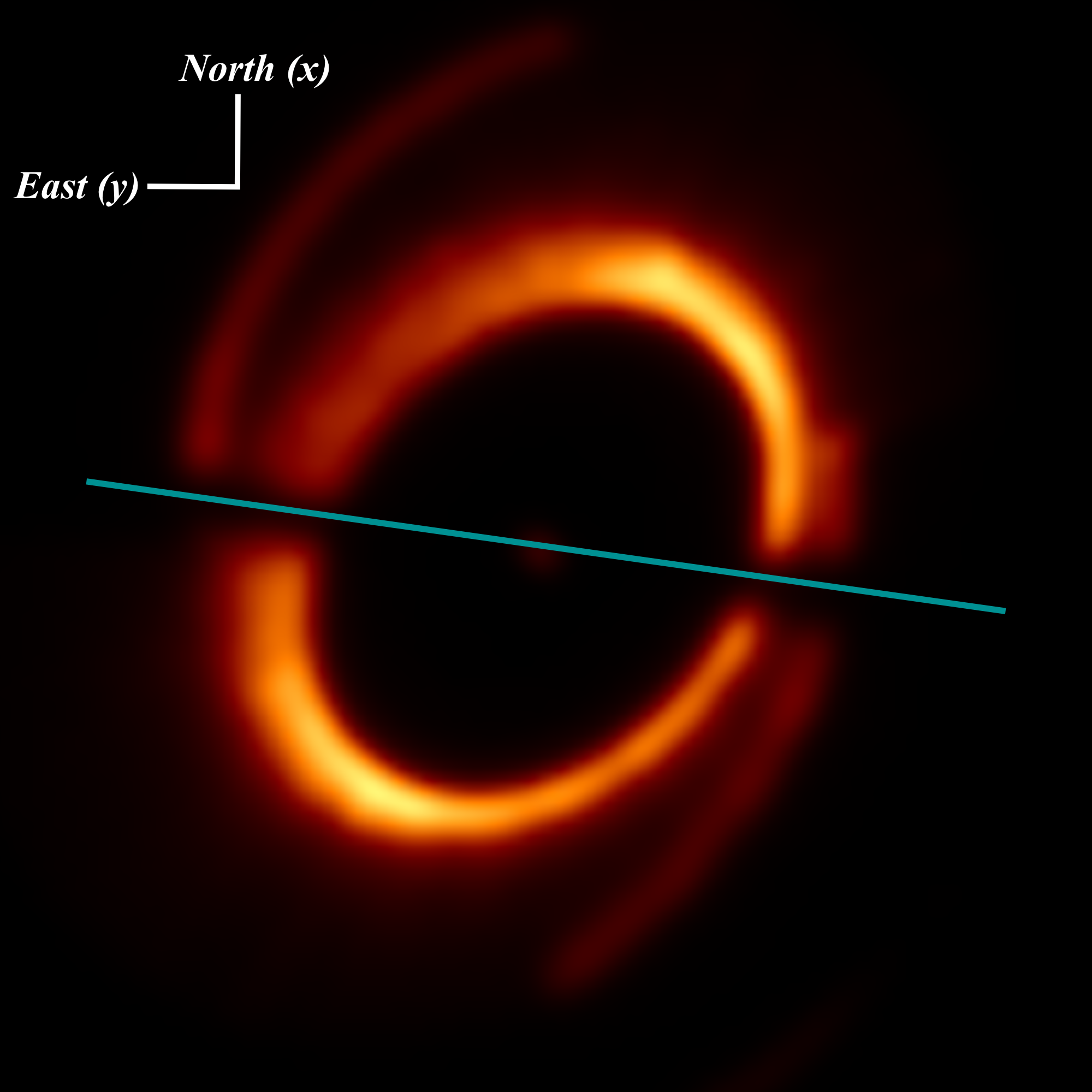}} &
\resizebox{0.33\hsize}{!}{\includegraphics{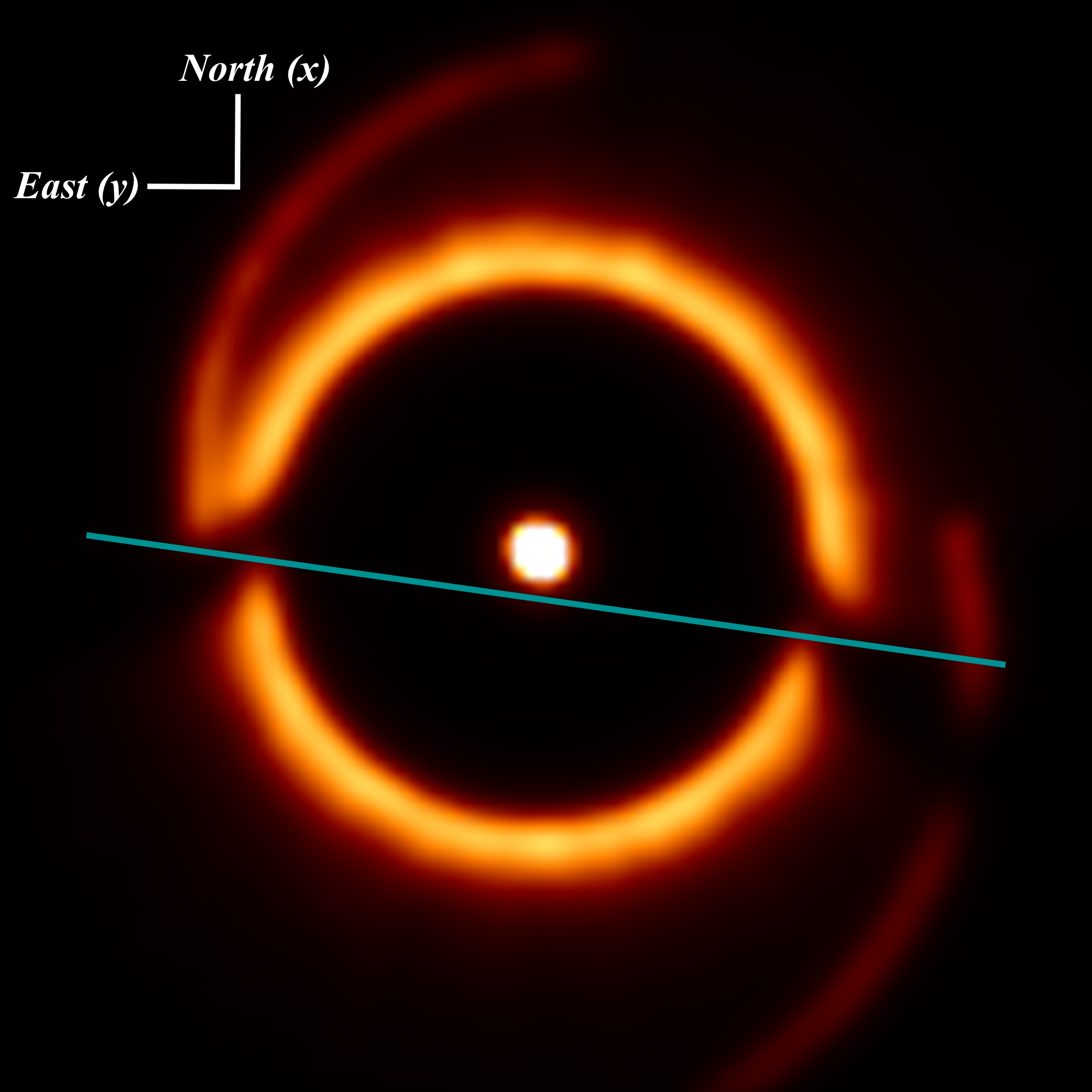}} &
\resizebox{0.33\hsize}{!}{\includegraphics{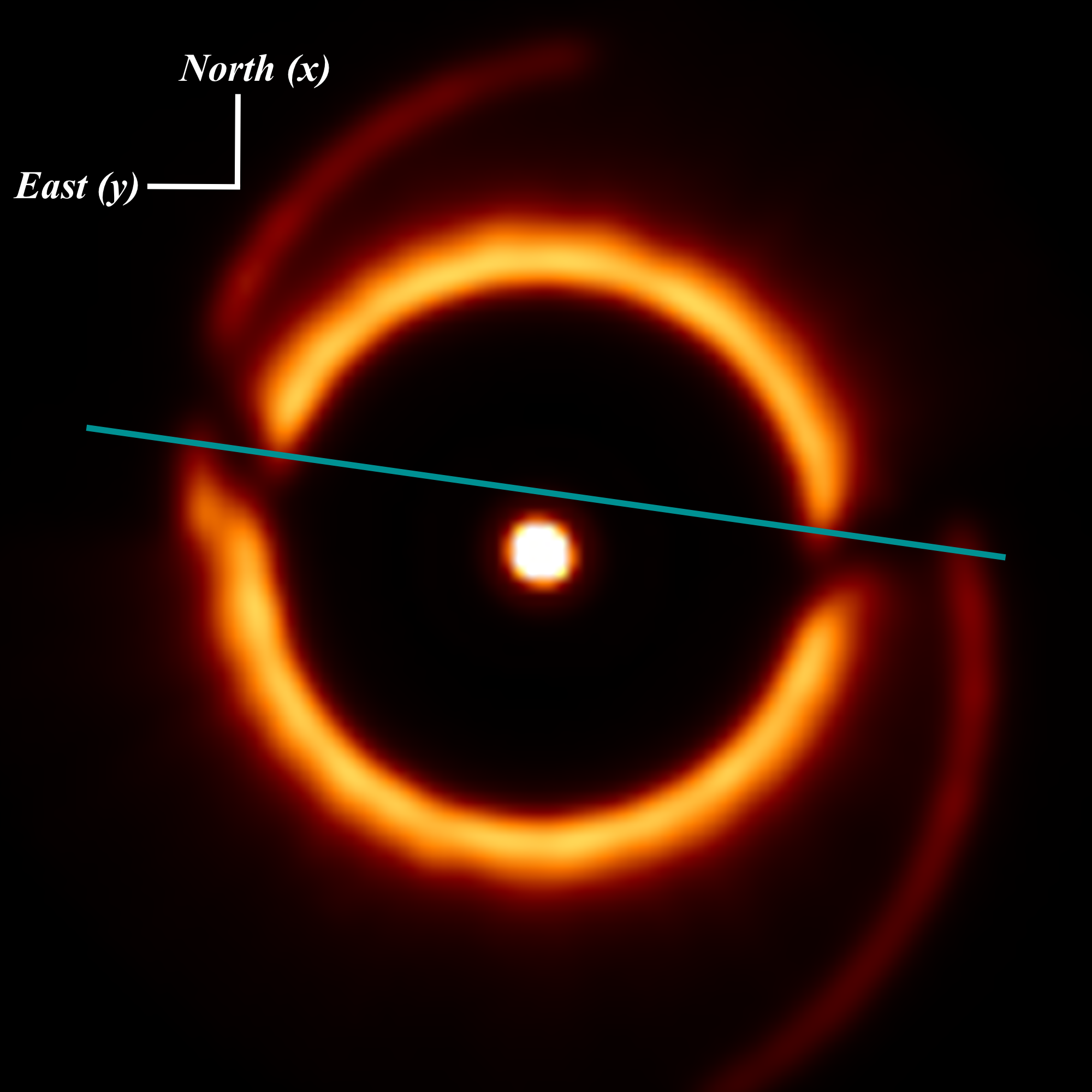}} &
\resizebox{0.33\hsize}{!}{\includegraphics{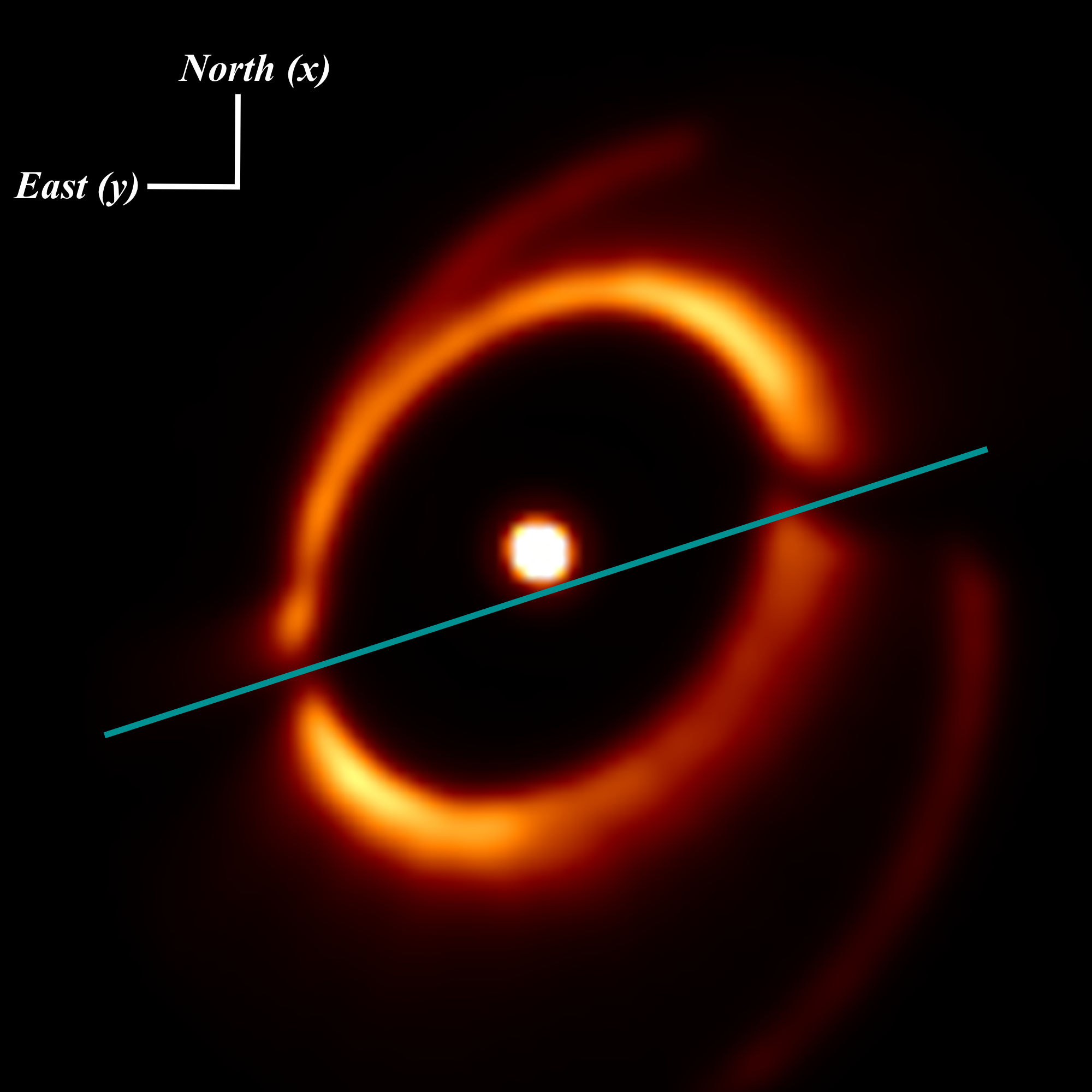}} \\
\end{tabular}}\end{center}
\caption{Simulated R-band images for the SPHERE instrument on the VLT. The model for the inner and outer disk is the same in all images but their geometric angles differ (given above each image). The blue line represents the line defined by Eqs.~(\ref{eq:posangle}) and (\ref{eq:offset}) with $h=4\,$au. The upper left model represents the best fit for the HD~100453 SPHERE image (see section~\ref{sec:HD100453}).}
\label{fig:models}
\end{figure*}

For a cartoon raytracing of the geometry see Fig.~\ref{fig:cartoon}. This image was generated with the three-dimensional (3D) version of the radiative transfer code MCMax \citep{2009A&A...497..155M} using the special `cartoon-mode'. In this mode the density distribution of the disk has a hard-edge surface instead of an exponential vertical density profile. This makes it much easier to visualise the geometry of the disks. As a basis we use the model created for the Herbig star HD~100453 presented in \citet{2017A&A...597A..42B}. We removed the spiral arms from this model and increased the size of the inner disk to make it clearly visible in the cartoon image. We note that the shadow in this cartoon representation shows the shadow on the East side as a hook due to the fact that we can see the spatially resolved shadow directly on the vertical wall. On the West side we see the shadow only on the surface of the disk due to the geometry of the system. In an observation with finite spatial resolution we expect the shadow on the surface to dominate on both sides.

\subsection{Position angle of the line connecting the shadows}

The intersection line of the two planes of the two disks defines the position angle of the line connecting the two shadows. We note that the position angle of the line connecting the shadows is independent of the height of the outer disk. The intersecting line of two planes needs to be parallel to both normal vectors and thus is defined by $\mathbf{a}=\hat{n}_1\times\hat{n}_2$.
\begin{equation}
\mathbf{a} = \begin{bmatrix}
        \sin(\theta_1)\cos(\theta_2)\cos(\phi_1)-\cos(\theta_1)\sin(\theta_2)\cos(\phi_2)       \\[0.3em]
        \sin(\theta_1)\cos(\theta_2)\sin(\phi_1)-\cos(\theta_1)\sin(\theta_2)\sin(\phi_2)       \\[0.3em]
        \sin(\theta_1)\sin(\theta_2)\sin(\phi_2-\phi_1)
     \end{bmatrix}.
\end{equation}
The position angle of the shadows, $\alpha$, is given by:
\begin{eqnarray}
\label{eq:posangle}
\alpha&=&\tan^{-1}\left(\frac{a_y}{a_x}\right)\\
&=&\tan^{-1}\left(\frac{\sin(\theta_1)\cos(\theta_2)\sin(\phi_1)-\cos(\theta_1)\sin(\theta_2)\sin(\phi_2)}{\sin(\theta_1)\cos(\theta_2)\cos(\phi_1)-\cos(\theta_1)\sin(\theta_2)\cos(\phi_2)}\right). \nonumber
\end{eqnarray}
This simplifies to $\alpha=\phi_1$ in the case where the outer disk is face on (i.e. $\theta_2=0$) or when $\phi_1=\phi_2$. 

\subsection{Offset with respect to the central star}

Let us define a point $\mathbf{x}=(x,y,z)$ in 3D space. All points $\mathbf{x}$ for which we have
\begin{equation}
\label{eq:disk1}
\hat{n}_1\cdot \mathbf{x}=0 \\
,\end{equation}
are in the plane of the inner disk. All points for which the equation
\begin{equation}
\label{eq:disk2}
\hat{n}_2\cdot \mathbf{x}=h
\end{equation}
holds are in the plane of the outer disk lifted $h$ above the disk midplane.
The line connecting the two shadows (the blue line in Fig.~\ref{fig:cartoon}) is given by all points for which both Eq.~(\ref{eq:disk1}) and (\ref{eq:disk2}) hold.
We already know the position angle of this line. We can easily compute the projected offset Northwards of the central star by setting $y=0$, giving, after some algebra,
\begin{equation}
\label{eq:offset}
x=\frac{h\cos(\theta_1)}{\cos(\theta_2)\sin(\theta_1)\sin(\phi_1)-\cos(\theta_1)\sin(\theta_2)\sin(\phi_2)}
.\end{equation}

\subsection{Application to simulated images}

We construct simulated polarimetric images for the SPHERE instrument on the VLT in the R-band. For this we use as a basis the model for the Herbig star HD~100453 from \cite{2017A&A...597A..42B} and rotate the inner and outer disk in different positions. We perform radiative transfer modelling using the 3D version of MCMax and convolve the resulting polarimetric images with the PSF of an $8\,$m class telescope and a seeing of $0.01''$. The resulting images are shown in Fig.~\ref{fig:models}. In these images we also plot the line defined by Eqs.~(\ref{eq:posangle}) and (\ref{eq:offset}) with $h=4\,$au, which roughly corresponds to the $\tau=1$ height in the R-band at $22\,$au for this model. It can be seen that for all orientations of the inner and outer disk the line connects the shadows in the images very well.

\begin{figure*}[!tp]
\centerline{\resizebox{0.8\hsize}{!}{\resizebox{0.5\hsize}{!}{\includegraphics{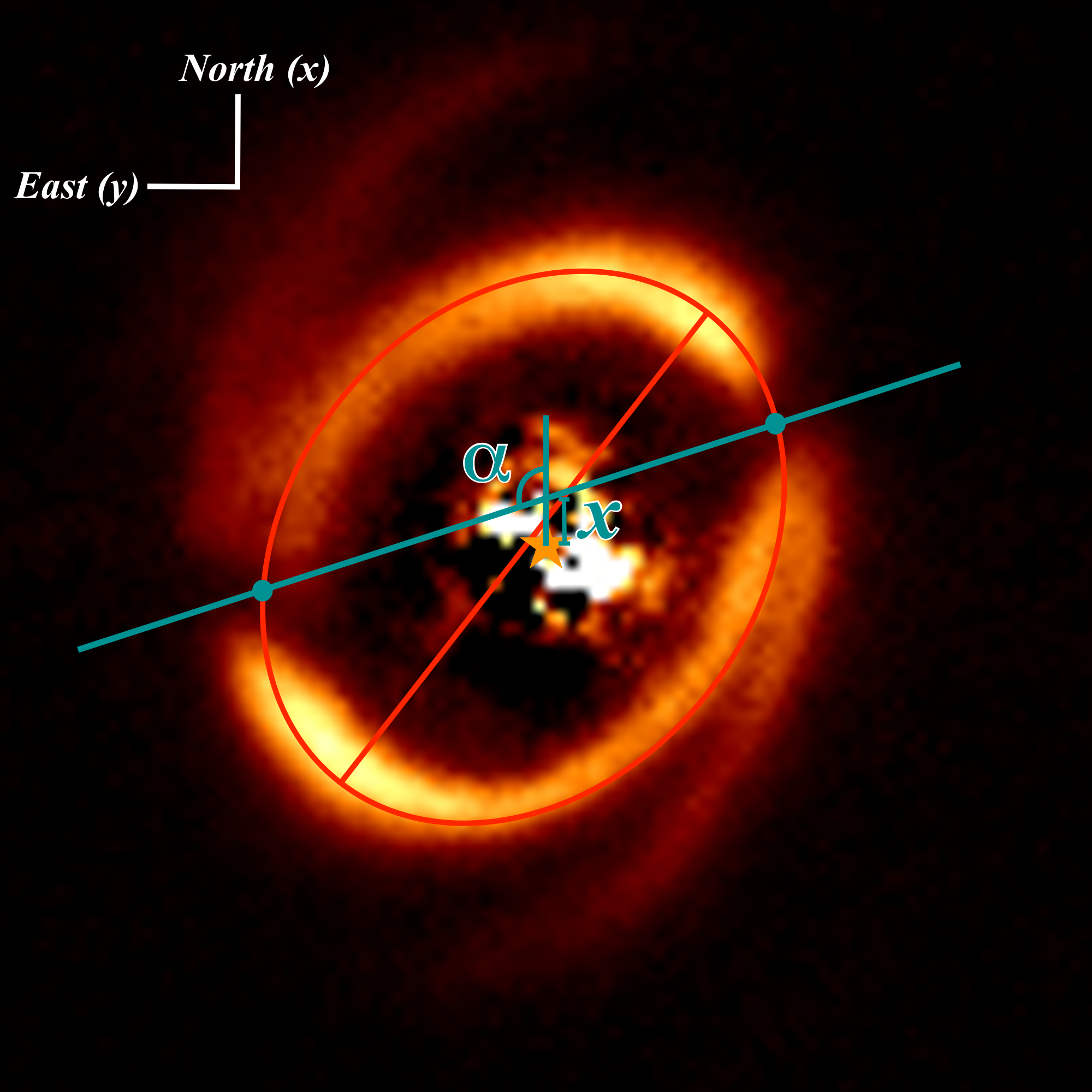}}\quad\resizebox{0.5\hsize}{!}{\includegraphics{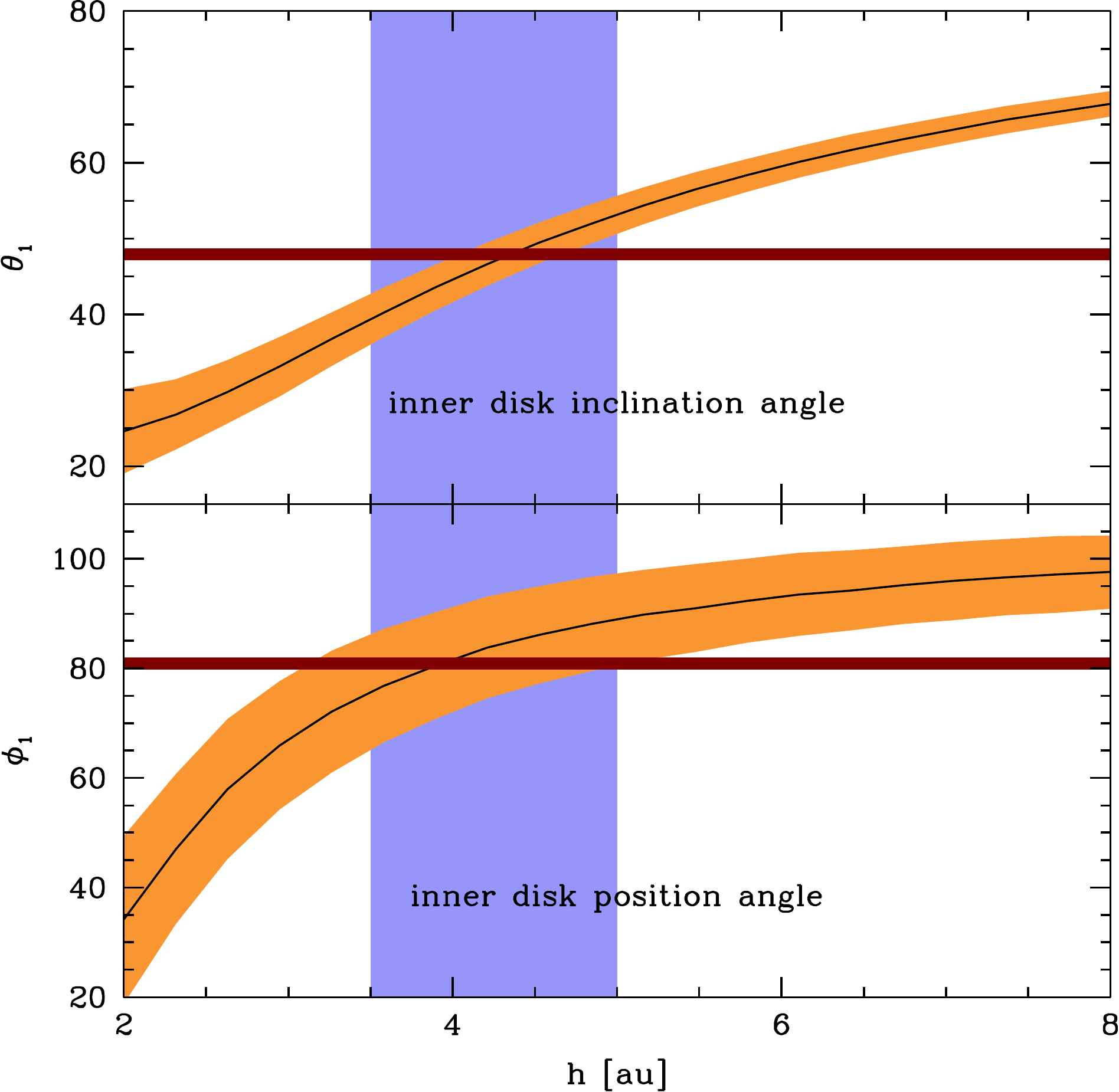}}}}
\caption{\emph{left:} SPHERE image of the disk surrounding HD~100453. The image presents the polarimetric intensity in the R-band. Indicated are also the ellipse of the outer disk \citep[from][]{2017A&A...597A..42B} and the position angle of the line connecting the shadows. \emph{right:} The derived inclination and position angle for the inner disk surrounding HD~100453 as a function of the assumed height of the scattering surface on the outer disk, $h$. The purple coloured regions indicate where we expect the height of the outer disk to be using models of the spectral energy distribution. The orange regions indicate the values for $\theta_1$ and $\phi_1$ given the uncertainties as defined in the text. The red horizontal bars indicate the values for $\theta_1$ and $\phi_1$ (with error bars) as derived from interferometric observations \citep{2017A&A...599A..85L}.}
\label{fig:HD100453sphere}
\end{figure*}

\section{Reconstructing inner disk geometries}
\label{sec:HD100453}

Now we will use the equations derived in the previous section to analyse the shadow locations in high contrast images of transition disks. Transition disks typically show a resolved cavity edge in scattered light which, assuming an intrinsic circular geometry, can be used to constrain the inclination, $\theta_2$ and position, $\phi_2$, angles with ellipse fitting. The position angle and the offset Northwards of the line connecting the shadows can also be derived from the image. Using Eqs.~(\ref{eq:posangle}) and (\ref{eq:offset}) we can then constrain the parameters of the inner disk. There is an intrinsic degeneracy in this procedure as we have two parameters we can derive from the image $(\alpha, x)$ and three free parameters to derive $(\theta_1, \phi_1, h)$. However, the height of the outer disk, $h$, can usually be roughly constrained from the spectral energy distribution.
We note that the observations we use here are performed in polarised intensity. Though for the location of the shadows this is not relevant, caution has to be taken locating the shadows when the degree of polarisation is highly variable over the image. This is especially the case for highly inclined disks. For the case we consider here, the shadows are too extreme to be caused by a decrease of the degree of polarisation alone.

\subsection*{HD~100453}

The transitional disk surrounding the Herbig star HD~100453 was imaged using the polarimetric mode of SPHERE. The images are presented in \cite{2017A&A...597A..42B} and shown in the left panel of Fig.~\ref{fig:HD100453sphere}. We measure the parameters $\alpha$ and $x$ from the image using the lines as shown in Fig.~\ref{fig:HD100453sphere}. When measuring the values of $\alpha$ and $x$, we have to consider that the equations for the connecting line were derived by considering the shadow as cast on the upper surface of the disk. We take the ellipse which fits the brightest parts of the disk image and connect the two points at the central position of the shadows along this ellipse. We find $\alpha=108^\circ$ and $x=0.031''$. Using a distance to the source of 114 parsec this implies $x=3.53\,$au. For the orientation of the outer disk we use the parameters as given in \cite{2017A&A...597A..42B} i.e. $\theta_2=-38^\circ$ and $\phi_2=142^\circ$. It is not feasible to quantify a reliable and meaningful statistical error on these values as they depend not on observational noise, but predominantly on the interpretation of the image and where one would locate the centre of the shadows. To make it clear how errors on these parameters influence the derived geometry of the inner disk, we use the rather arbitrary values of $5^\circ$ on $\theta_2, \phi_2$ and $\alpha$ and for $x$ we take an error of $5\%$. We can now use Eqs.~(\ref{eq:posangle}) and (\ref{eq:offset}) to derive the orientation of the inner disk as a function of the height of the outer disk. This is plotted in the right panel of Fig.~\ref{fig:HD100453sphere}. From the modelling presented in \cite{2017A&A...597A..42B} we found that the scattering surface of the inner edge of the outer disk lies somewhere in the range $3.5\,$au$<h<5.0\,$au. This region is coloured purple in the right panel of Fig.~\ref{fig:HD100453sphere}. Also indicated are the angles as derived from interferometric observations by \cite{2017A&A...599A..85L}. We see that the orientation of the inner disk as derived from the positioning of the shadows on the outer disk is roughly consistent with the interferometric measurements. This was already concluded from the very good agreement between observed and modelled images in \cite{2017A&A...597A..42B}. Although the exact combination $(\theta_1,\phi_1)=(47.9^\circ,81)\pm(0.8^\circ,1^\circ)$ derived from interferometry gives $\alpha=105.7^\circ$, a slight adjustment towards e.g. $(\theta_1,\phi_1)=(45^\circ,82^\circ)$ gives exactly $\alpha=108^\circ$. This conclusion seems to contrast statements in the paper by \cite{2017ApJ...838...62L}. However, as is explained in the accompanying erratum of that paper (Long et al. in prep), there was a misinterpretation of the angles in the radiative transfer code used, and the angles quoted in the original paper are incorrect. The actual angles used in the radiative transfer modelling indeed nicely agree with those given in \cite{2017A&A...597A..42B}, \cite{2017A&A...599A..85L} and in this paper.
The upper left panel of Fig.~\ref{fig:models} presents the model image for HD~100453 according to \cite{2017A&A...597A..42B} (with the angles slightly adjusted according to the findings of this paper). We can see that the position angle of the line connecting the shadows is perfectly reproduced.


\section{Conclusions}

We present an analytical framework that can be used to derive the orientation of the inner disk of misaligned transitional disk systems by using the location of shadow features. We show that the position angle of the connecting line between the shadows is generally very different from the position angle of the inner disk. This somewhat counterintuitive result has to be kept in mind for correct interpretation of high contrast images. We outline a methodology that includes estimating the height of the outer disk from SED modelling to constrain the inner disk inclination and position angles.


\bibliographystyle{aa}
\bibliography{biblio}

\end{document}